\documentclass[twocolumn]{aastex631}
\usepackage{bm}
\usepackage{amsmath}
\usepackage{booktabs}
\usepackage{tabularx}
\usepackage{array}

\begin{document}
\title{Supernova Classification using the Recurrent Neural Network in the CSST Ultra-Deep Field Survey}

\correspondingauthor{Yan Gong}
\email{Email: gongyan@bao.ac.cn}

\author{Minglin Wang}
\affiliation{National Astronomical Observatories, Chinese Academy of Sciences, Beijing 100101, People's Republic of China}
\affiliation{University of Chinese Academy of Sciences, Beijing 100049, China}   

\author{Yan Gong*}
\affiliation{National Astronomical Observatories, Chinese Academy of Sciences, Beijing 100101, People's Republic of China}
\affiliation{University of Chinese Academy of Sciences, Beijing 100049, China}      
\affiliation{Science Center for China Space Station Telescope, National Astronomical Observatories, Chinese Academy of Sciences, 20A Datun Road, Beijing 100101, China}

\author{Dejia Zhou}
\affiliation{National Astronomical Observatories, Chinese Academy of Sciences, Beijing 100101, People's Republic of China}
\affiliation{University of Chinese Academy of Sciences, Beijing 100049, China}   

\author{Xuelei Chen}
\affiliation{National Astronomical Observatories, Chinese Academy of Sciences, Beijing 100101, People's Republic of China}
\affiliation{University of Chinese Academy of Sciences, Beijing 100049, China}      
\affiliation{Center for High Energy Physics, Peking University, Beijing 100871, China}
\affiliation{Department of Physics, College of Sciences, Northeastern University, Shenyang 110819, China}
\affiliation{State Key Laboratory of Radio Astronomy and Technology, China}



\begin{abstract}
{We study supernova (SN) classification using the Recurrent Neural Networks (RNNs) within the Chinese Space-station Survey Telescope Ultra-Deep Field (CSST-UDF) photometric survey and explore the improvements of the cosmological constraint.} {We simulate Type Ia supernovae (SNe~Ia) and core collapse supernovae (CCSNe) using SNCosmo with SALT3 SN~Ia model and CCSN templates, and apply the SuperNNova (SNN) program for classification.} Our study indicates that the SNN combined with the Joint Light-curve Analysis cuts can enhance the purity of the CSST-UDF SN~Ia sample up to over 99.5\% with 2,193 SNe~Ia and 4 CCSNe, {which can significantly increase the reliability of the cosmological constraints. The method based on the Bayesian Estimation Applied to Multiple Species with Bias Corrections framework is used to correct the SN~Ia magnitude bias caused by the selection effect and CCSN contamination, and the Markov Chain Monte Carlo (MCMC) method is employed for cosmological constraints.}
We find that the accuracy of the constraints on the matter density $\Omega_{\rm M}$ and {the equation of state of dark energy parameter $w$} can achieve 14\% and 18\%, respectively, assuming the flat $w$CDM model. {This result is comparable to current surveys relying on spectroscopic confirmation. Our results indicate that our data analysis method is effective, and the CSST-UDF SN photometric survey is a powerful tool to explore the expansion history of the Universe.}
\end{abstract}


\section{Introduction} \label{sec:intro}
Type Ia supernovae (SNe~Ia), which are regarded as standard candles in cosmology, reveal the accelerating expansion of the Universe and the possible existence of dark energy \citep{Riess98,Perlmutter_1999}. {Recently, many SN~Ia surveys have been performed (e.g., Dark Energy Survey, DES, \citealt{DES_3YEARS_SN,DES5Year23,DES_BIASFROMCLASS,descollaboration2024dark,DESSNchen2024evaluatingcosmologicalbiasesusing,des5yrNhz,DES_2024beyondLCDM,descollaboration2025darkenergysurveyimplications}; Supernova H0 for the Equation of State, SH0ES, \citealt{SH0ESRiess_2019,SH0ESBreuval_2024}; and Pantheon+, \citealt{pahtheondataset,PANTHEONCOSMOBrout_2022})}, which have greatly promoted the study of the expansion history of the Universe. {However, most SNe~Ia in these samples are located at relatively low redshifts. To accurately investigate important cosmological problems, such as the evolution of the dark energy equation of state, high-redshift samples are needed. These samples allow us to measure the cosmic expansion history and distances across a broader redshift range.

Photometry surveys are an effective tool for detecting large numbers of high-redshift ($z>1$) SNe Ia}, especially in the era of Stage IV surveys like the Vera Rubin Observatory's Legacy Survey of Space and Time (LSST, \citealt{LSST_INTRO_2019ApJ...873..111I,LSSTSN_SIMU_Kumar_2025}), $\it Euclid$ \citep{EUCLIDOverview2025,EUCLI_AND_RUBINB_ailey_2023}, Nancy Grace Roman Space Telescope (RST, \citealt{RSTSN2021,RST2024SPIE13092E..0SS}), and the Chinese Space-station Survey Telescope (CSST, \citealt{zhan11,Gong19,HuZhan2021,gong2025,CSSTCOR2025}. 
For instance, the CSST is a space-based telescope under the China Manned Space Program, which is expected to be launched around  2027. The CSST is designed with a 2-meter primary mirror and a field of view of 1.1 square degrees. It can simultaneously perform the photometric and spectroscopic surveys in wide, deep, and ultra-deep fields, covering the wavelength range from about 2500\AA\ to 10000\AA. {The detector layout and filter bandpasses of CSST can be found in \citet{HuZhan2021} and \citet{wang2024}.}

{The CSST Ultra-Deep Field (CSST-UDF) survey is expected to observe a sky area of about 9 square degrees in the first two years after launch. The exact location of the field will be selected within a high Galactic latitude region to study high-redshift galaxies and SNe.
The selected field will be observed 60 times with a cadence of approximately 12 days, and each exposure lasting 250 seconds. For a single visit, the expected magnitude limits for point sources 5$\sigma$ detection of the seven filters, i.e. $\it NUV$, $\it u$, $\it g$, $\it r$, $\it i$, $\it z$, and $\it y$, are 25.3, 25.7, 26.4, 26.1, 25.9, 25.4, and 24.4 AB mag, respectively, and can reach 28.0, 28.0, 28.7, 28.4, 28.2, 27.7, and 27.1 AB mag for 60 exposures \citep{CAOYE22,gong2025,CSSTCOR2025}.} Therefore, it is expected that the CSST-UDF survey can accurately measure more than 2000 SN~Ia with a large fraction of the sample at high redshifts \citep{LISHIYU2023,wang2024}. {Furthermore, the CSST Wide Field survey covers a survey area of 17,500 deg$^2$, with 5$\sigma$ point source magnitude limits of 25.4, 25.4, 26.3, 26.0, 25.9, 25.2, and 24.4 AB mag for the $NUV$, $u$, $g$, $r$, $i$, $z$, and $y$ bands, respectively. This survey is capable of detecting a large number of SNe for cosmological studies \citep{Liu_2024}.}

However, like other photometric surveys, the CSST-UDF SN Ia survey faces significant challenges in supernova identification and classification. { The primary difficulty with large surveys like CSST-UDF is that the availability of spectroscopic follow-up resources is limited and expensive.} Thus, it is essential to develop methods that can effectively classify supernovae using only photometric data. Traditional photometric SN classification methods rely on template fitting and parameter constraints.
{However, since certain regions of the light-curve parameter space for some core collapse supernovae (CCSNe) may overlap with SNe~Ia~\citep{VINCENZI2021,VIN22,wang2024}, it is difficult for these methods to effectively separate such CCSNe from SNe~Ia. To address this limitation, we need approaches that can better extract the light-curve parametric features of different SNe subtypes.}  

{Machine learning, particularly RNNs, is highly suitable for SN classification because RNNs are naturally designed to handle sequential data with heterogeneous time sampling, which is a key characteristic of photometric light curves.} RNNs have a long development history, which can be traced back to the early neural network models. \citet{hopfield1982} introduced one of the earliest recurrent network models, \citet{Rumelhart1986LearningRB} developed the backpropagation algorithm, and \citet{elman1990finding} proposed a simplified RNN architecture that popularized their use in sequential data modeling. {The demonstrated performance of these architectures in processing time-series data makes them directly applicable to SN light-curve analysis. In particular, RNN-based methods enable SN type prediction even during early observational phases \citep{Leoni22}. In the cosmological constraint stage, we apply the Bayesian Estimation Applied to Multiple Species with Bias Correction (BBC) method proposed by \citet{KUNZ17BEAMS,KS17BBC}. This approach can effectively correct the biases introduced by selection effects and sample contamination.} 
In this study, we focus on studying SN classification in the CSST-UDF photometric survey by employing the SuperNNova \citep[SNN,][]{SuperNNova}, and explore the improvement of the cosmological constraints. 

The paper is organized as follows: 
in Section~\ref{sec:sn-sim}, we describe the simulation procedure of generating mock SN light curve data; 
in Section~\ref{sec:classification}, we present the classification model and corresponding results; 
in Section~\ref{sec:cosmology}, we discuss the cosmological constraint results obtained using a BBC-like framework;
we give the conclusions in Section~\ref{sec:conclusion}.

\begin{table*}
\centering
\caption{Supernova model parameters and volumetric rates ($10^{-4} \, \mathrm{yr}^{-1} \mathrm{Mpc}^{-3} h_{70}^3$) used in this work. Subtype fractions and Gaussian LFs are provided for SNe~Ia (rest-frame $B$-band; \citealt{SH0ESMABS-19.253RIESS}) and CCSNe (rest-frame $R$-band; \citealt{V19, J17}). Volumetric rates across redshift bins are adopted from \citet{SN_RATE14} for SNe~Ia and \citet{strolger15} for CCSNe.}
\label{tab:snpar}
\begin{tabular}{l c c c c c c c}
\toprule
\multicolumn{3}{c}{Subtype Characterization} & & \multicolumn{4}{c}{volumetric rates} \\
\cmidrule{1-3} \cmidrule{5-8}
SN Type & Fraction (\%) & LF ($\mu \pm \sigma$) [mag] & & $z_{\rm Ia}$ & $\text{Rate}_{\rm Ia}$ & $z_{\rm CCSN}$ & $\text{Rate}_{\rm CCSN}$ \\
\midrule
SN Ia   & ---  & $-19.25 \pm 0.10$ & & $0.0 - 0.5$ & 0.36 & $0.1 - 0.5$ & 2.13 \\
SN IIL  & 7.9  & $-18.28 \pm 0.45$ & & $0.5 - 1.0$ & 0.51 & $0.5 - 0.9$ & 3.86 \\
SN IIP  & 57.0 & $-16.67 \pm 1.08$ & & $1.0 - 1.5$ & 0.64 & $0.9 - 1.3$ & 3.07 \\
SN IIb  & 10.9 & $-16.69 \pm 1.99$ & & $1.5 - 2.0$ & 0.72 & $1.3 - 1.7$ & 3.25 \\
SN IIn  & 4.7  & $-17.66 \pm 1.08$ & & & & & \\
SN Ic   & 11.9 & $-17.44 \pm 0.66$ & & & & & \\
SN Ib   & 7.5  & $-18.26 \pm 0.15$ & & & & & \\
\bottomrule
\end{tabular}
\end{table*}

\section{Light Curve Generation} \label{sec:sn-sim}

\begin{figure*}
  \centering
  \begin{minipage}[b]{0.48\textwidth}
    \centering
    \includegraphics[width=\textwidth]{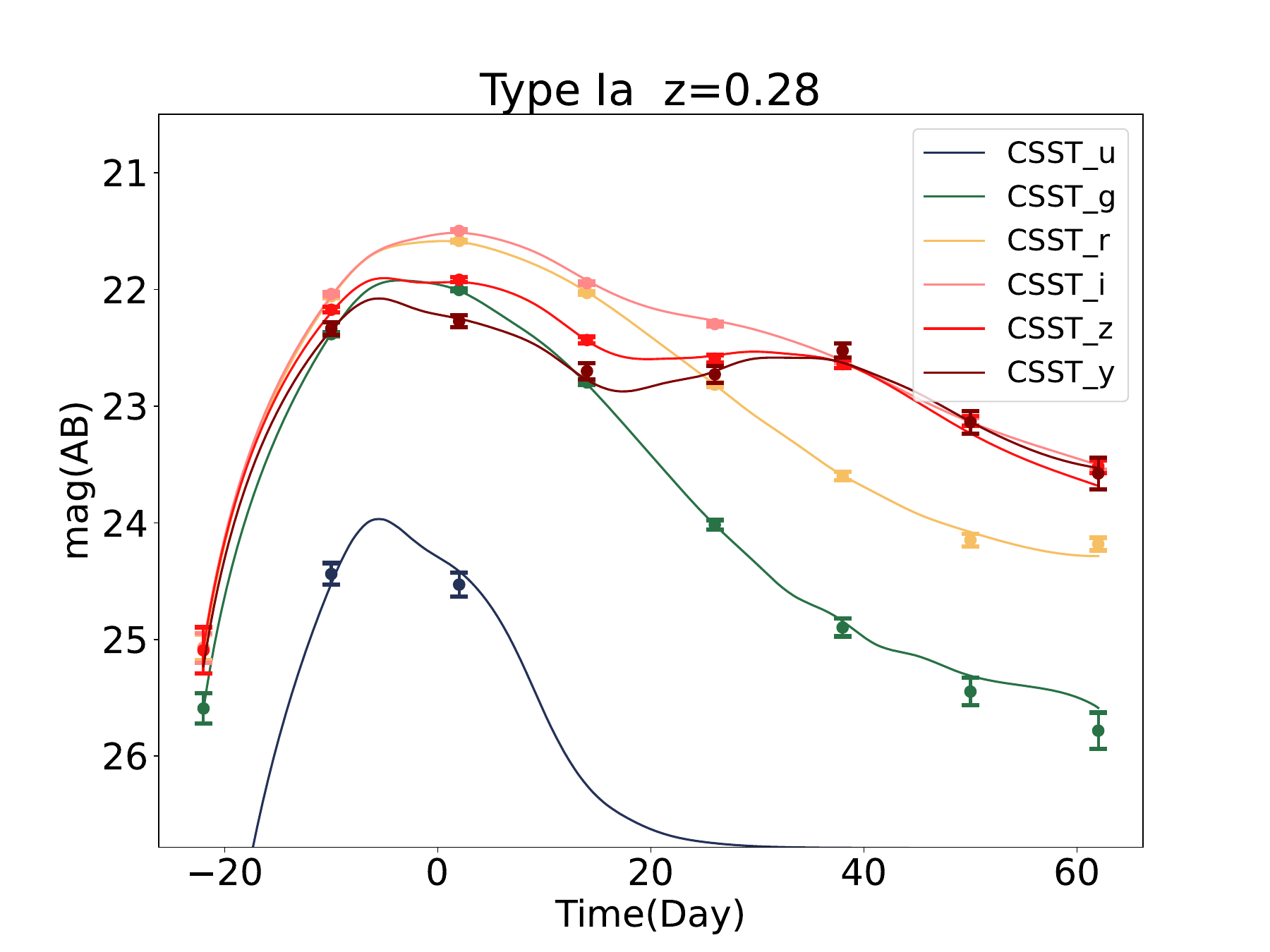}
  \end{minipage}
  \hfill
  \begin{minipage}[b]{0.48\textwidth}
    \includegraphics[width=\textwidth]{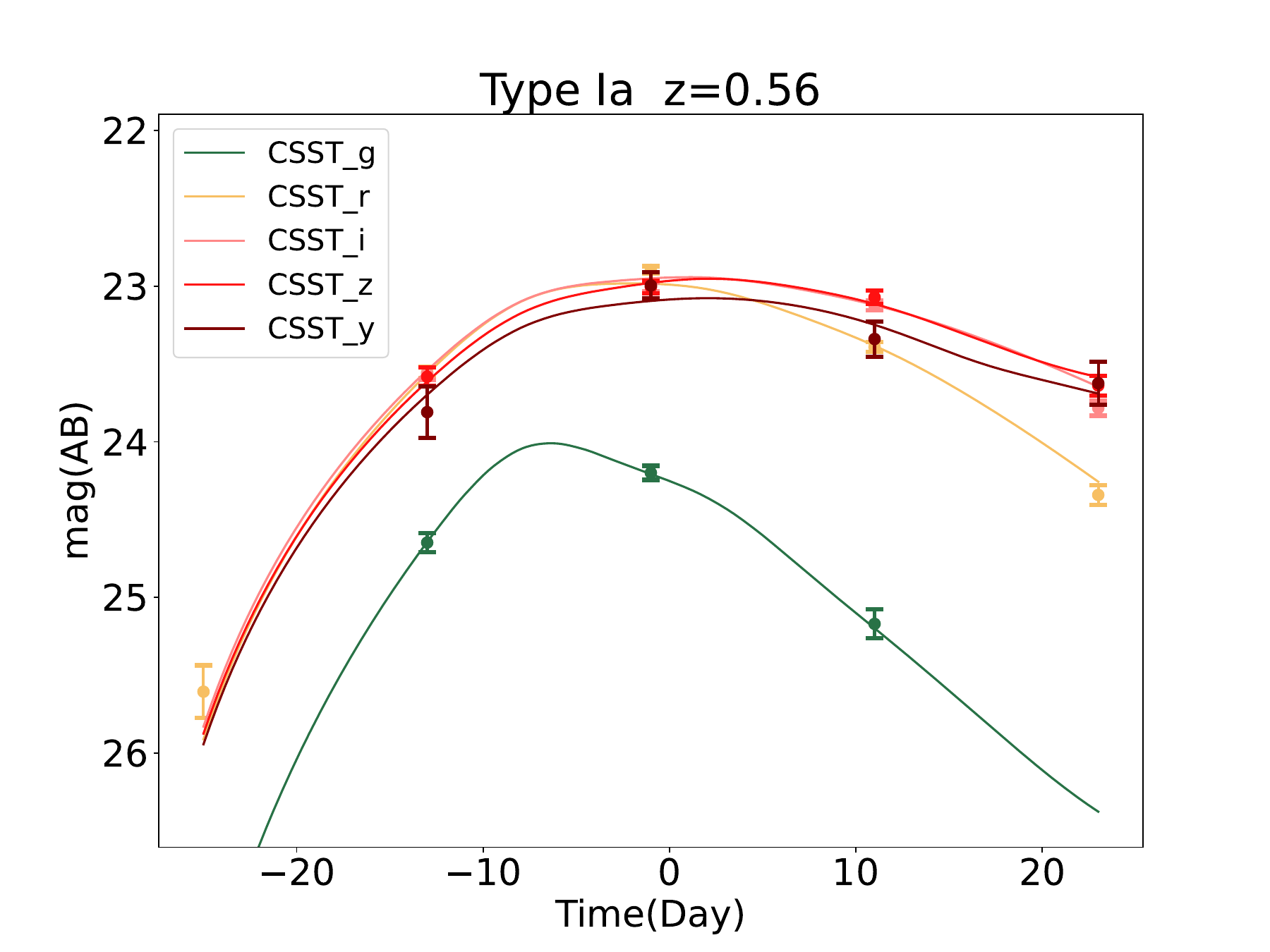}
  \end{minipage}
\vspace{0.5\baselineskip}
  \begin{minipage}[b]{0.48\textwidth}  

      \centering
      \includegraphics[width=1\linewidth]{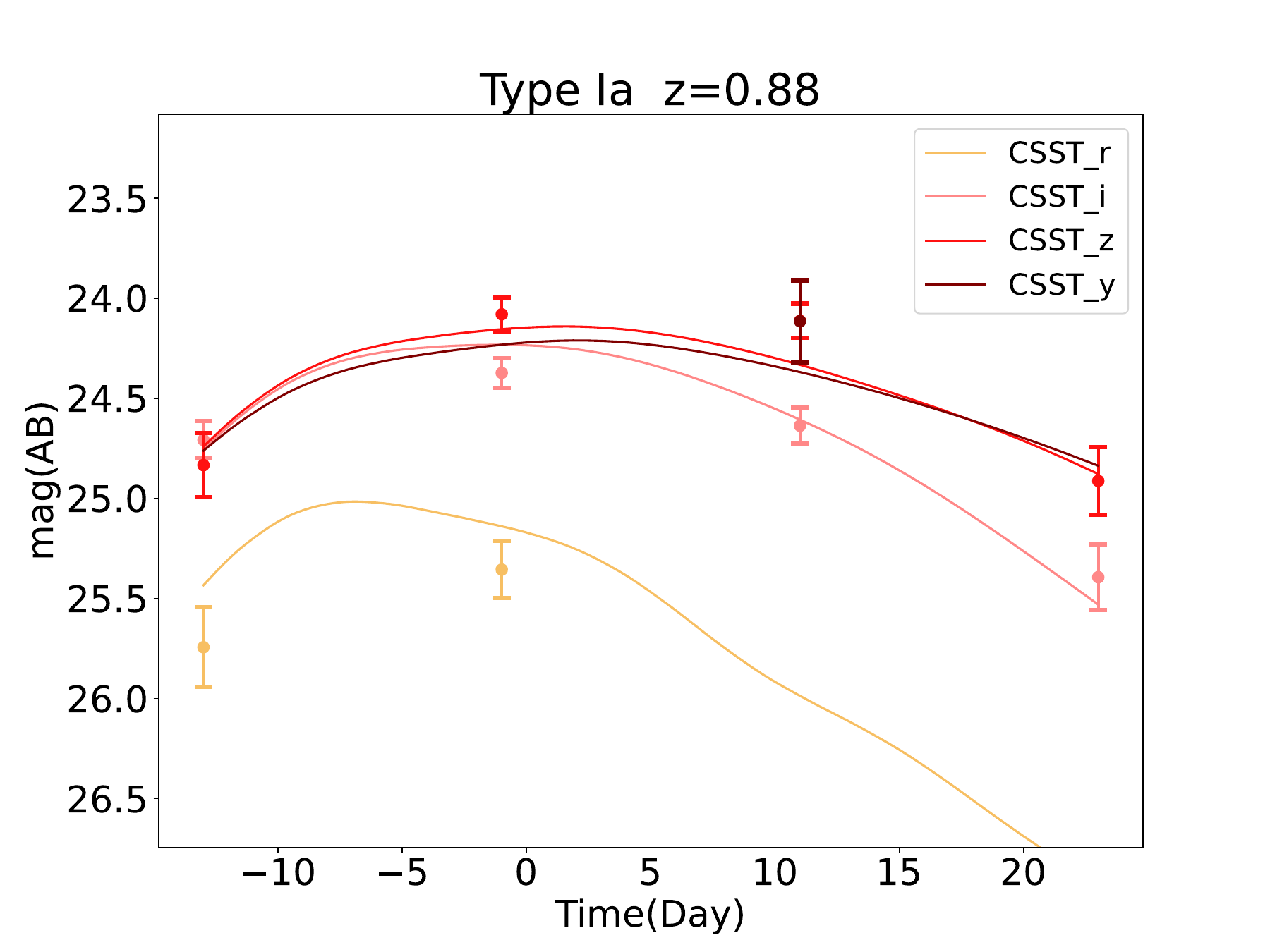}
\end{minipage}
  \hfill
  \begin{minipage}[b]{0.48\textwidth}
        \includegraphics[width=1\linewidth]{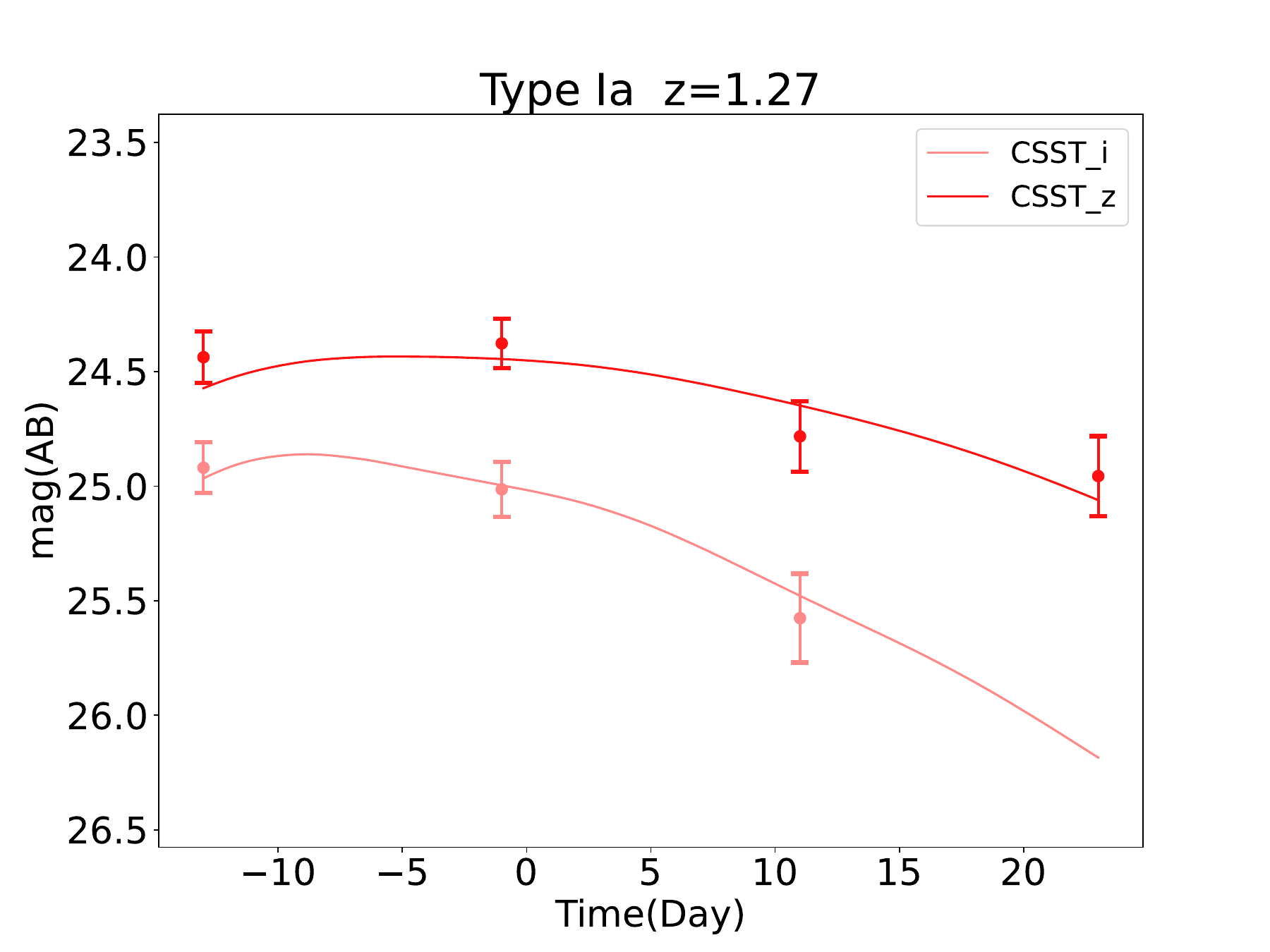}
        \end{minipage}
 \caption{  \label{fig:Iasnlight curve}The mock light curve examples for SNe~Ia in different CSST-UDF photometric bands at redshifts between $z = 0.28$ and 1.3. The data points with error bars represent the simulated observational fluxes and their associated uncertainties. The solid lines indicate the theoretical light curves derived from individual fiducial parameters of each SN.}
\end{figure*}

\begin{figure*}
    \hfill
    \centering
    \vspace{0.5\baselineskip}
  \begin{minipage}[b]{0.48\textwidth}
       \includegraphics[width=1\linewidth]{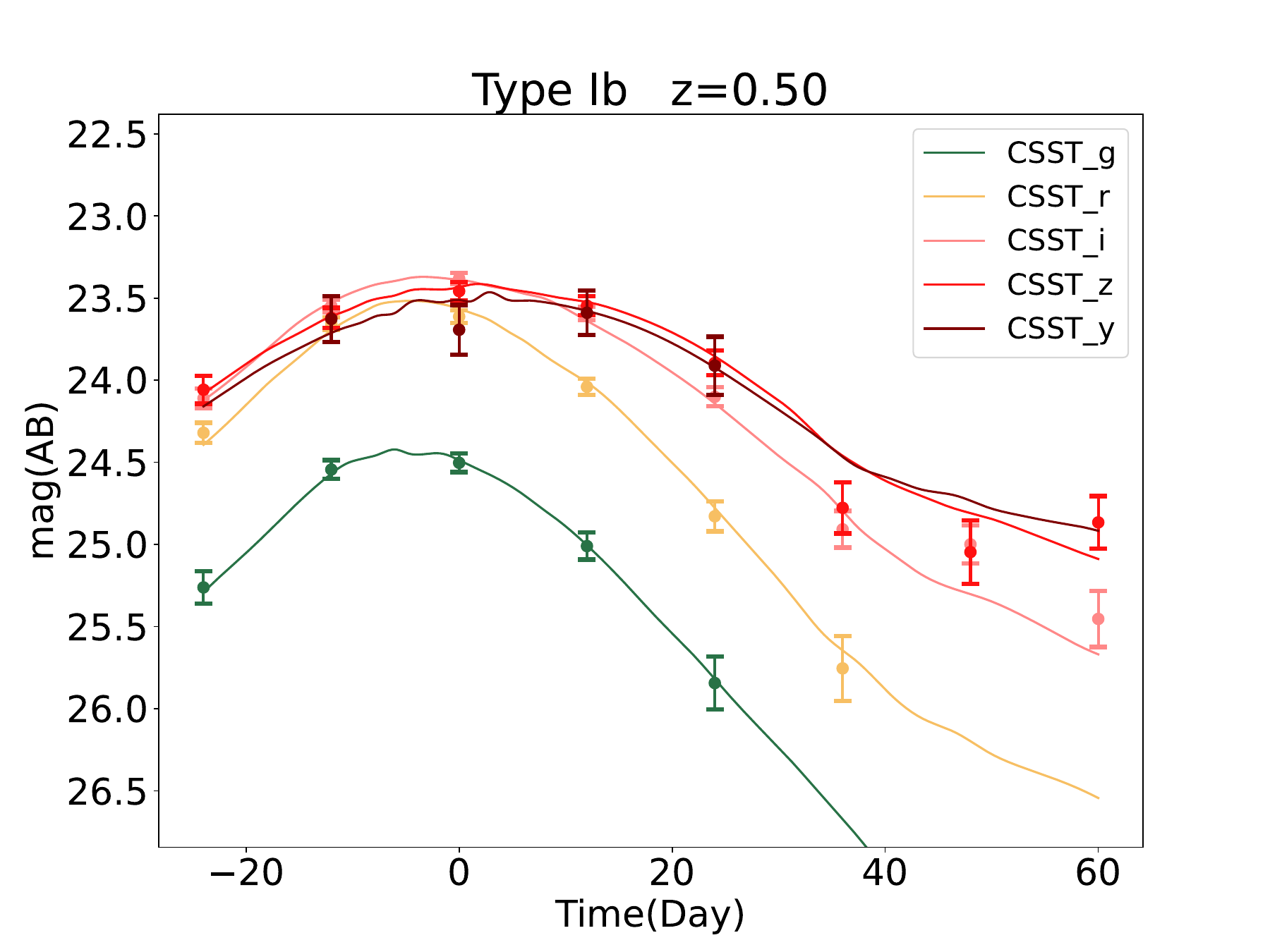}
     \end{minipage}    
    \hfill
    \vspace{0.5\baselineskip}
  \begin{minipage}[b]{0.48\textwidth}
      \includegraphics[width=1\linewidth]{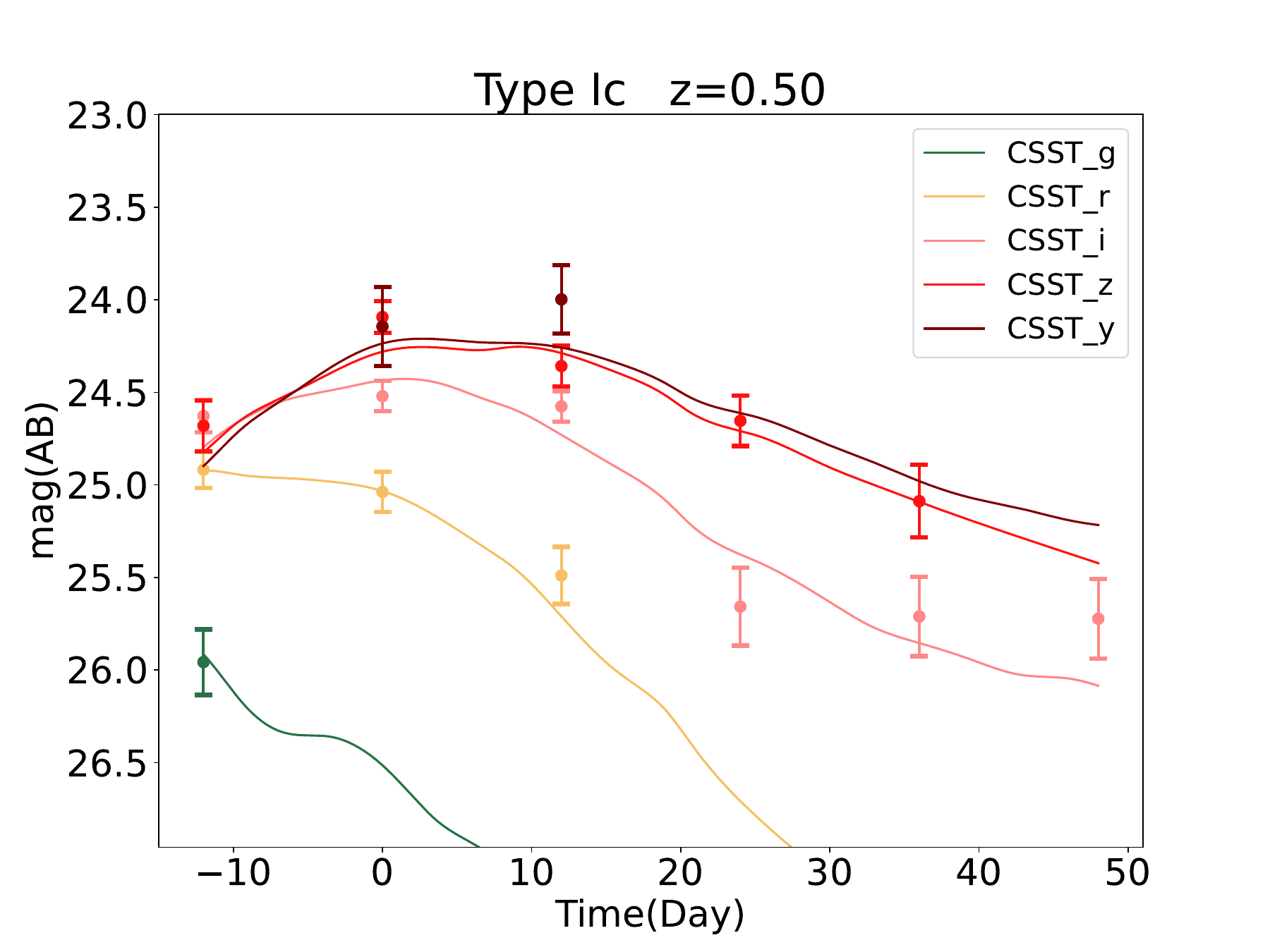}
    \end{minipage}
   \hfill
    \centering
    \vspace{0.5\baselineskip}
  \begin{minipage}[b]{0.48\textwidth}
       \includegraphics[width=1\linewidth]{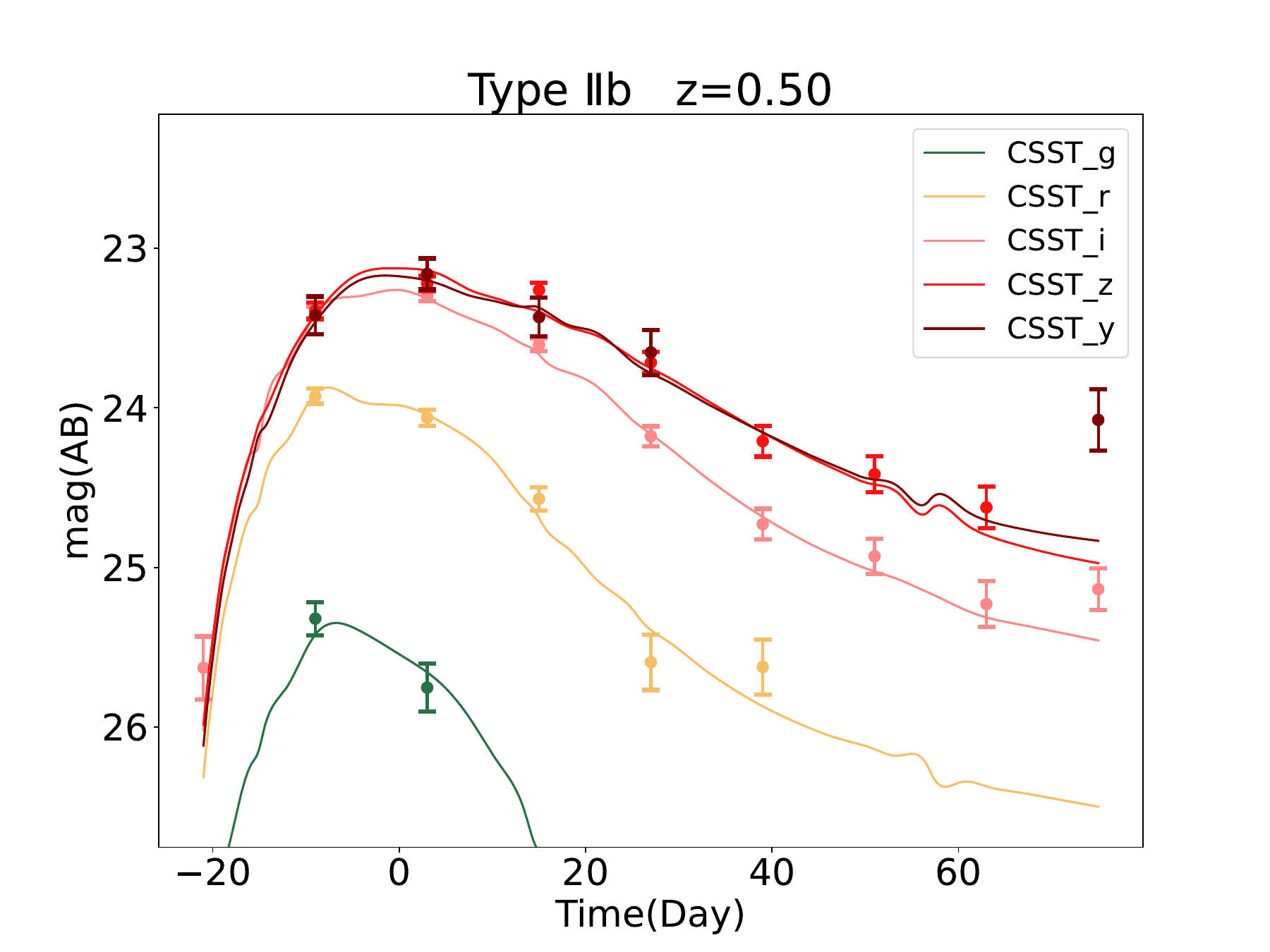}
     \end{minipage}    
    \hfill
    \vspace{0.5\baselineskip}
  \begin{minipage}[b]{0.48\textwidth}
      \includegraphics[width=1\linewidth]{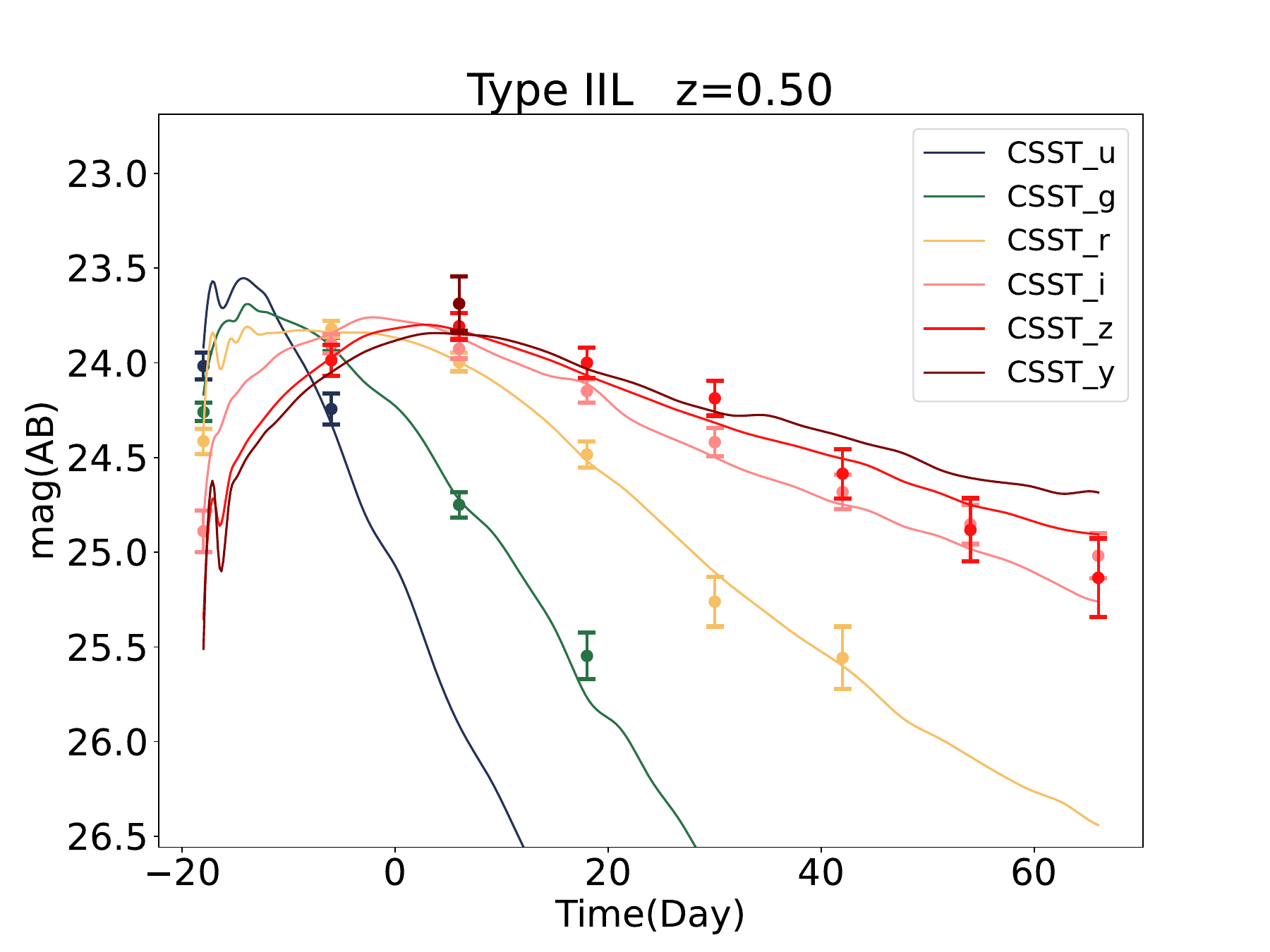}
    \end{minipage}
   \hfill
    \centering
    \vspace{0.5\baselineskip}
  \begin{minipage}[b]{0.48\textwidth}
       \includegraphics[width=1\linewidth]{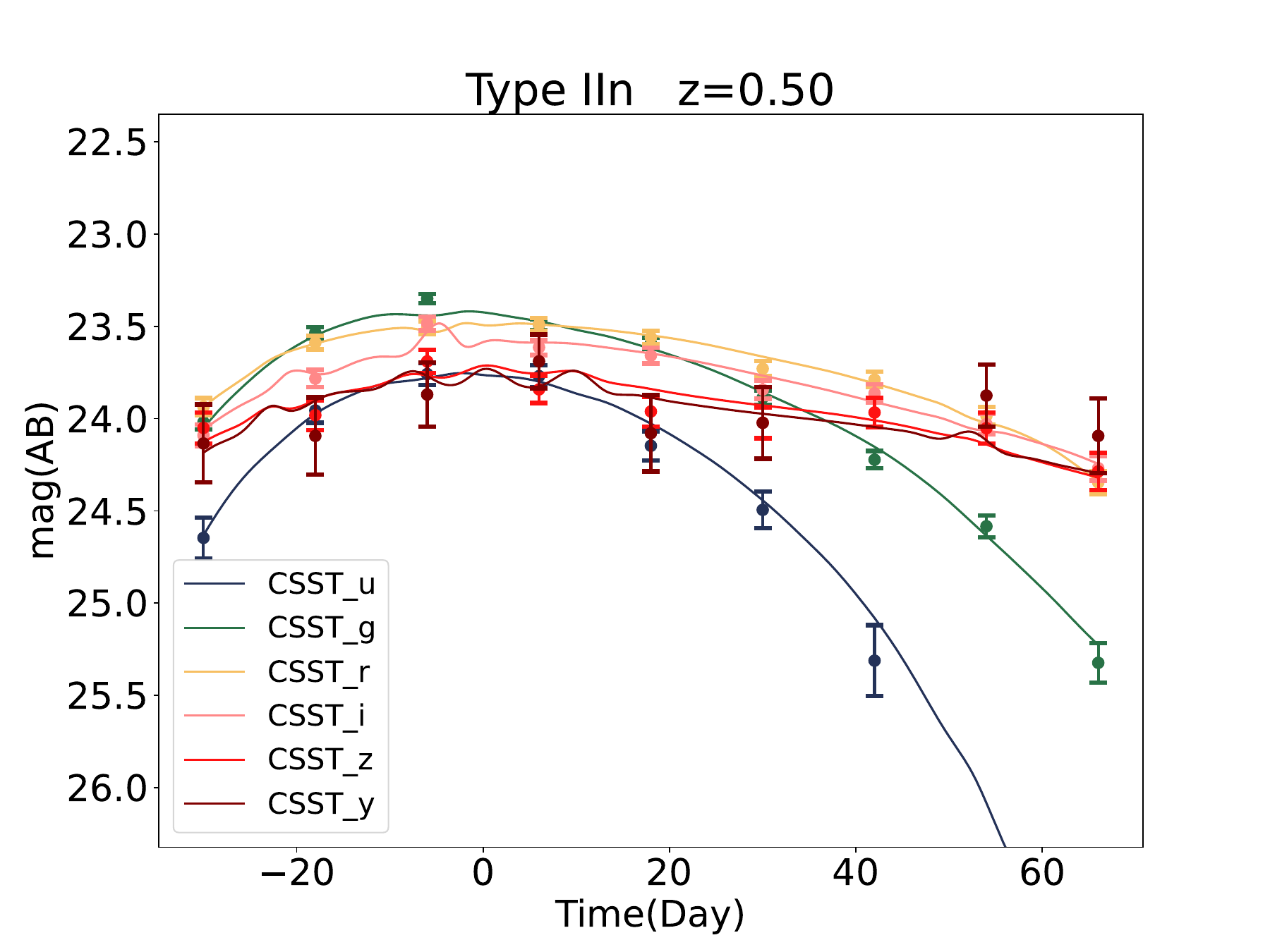}
     \end{minipage}    
    \hfill
    \vspace{0.5\baselineskip}
  \begin{minipage}[b]{0.48\textwidth}
      \includegraphics[width=1\linewidth]{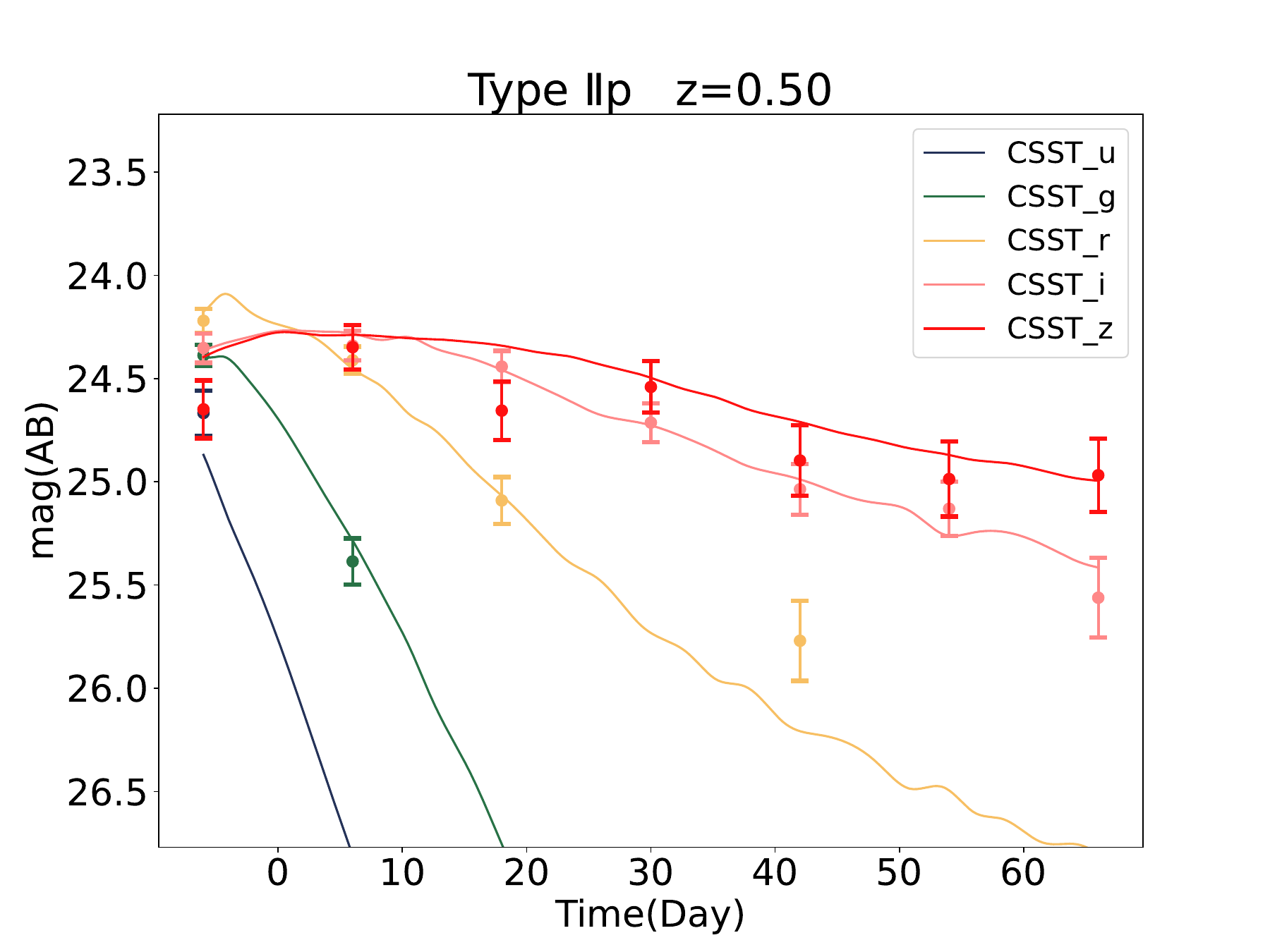}
    \end{minipage}
 \caption{\label{fig:ccsnlight curve}The mock light curve examples at $z \simeq 0.5$ for the six types of CCSNe considered in this study. The data points with error bars represent the simulated observational fluxes and their associated uncertainties. The solid lines indicate the theoretical light curves derived from individual fiducial parameters of each SN.}
\end{figure*}

{We employ \texttt{SNCosmo} \citep{SNCOSMO} as the basic framework for generating mock light curves and performing light-curve fitting. 
The training and testing datasets are simulated using the SALT3 model \citep{SALT3_Kenworthy2021} and CCSN templates \citep{V19}. The SN light curves are first generated based on the SN parameters, the CSST instrumental design and the CSST-UDF survey strategy. Then selection criteria are applied to filter SNe based on the signal-to-noise ratio (SNR) of the light curve mock data. For the mock observations, we adopt a fiducial flat $w$CDM cosmological model with $\Omega_{\rm M} = 0.3$, $w = -1$, and $h=0.7$.}

{The SN parameters adopted in this study follow \citet{wang2024}. The specific simulation parameters, including SN volumetric rates, luminosity functions (LFs), and CCSN subtype fractions, are detailed in Table~\ref{tab:snpar}. For SNe~Ia, additional parameters such as stretch ($x_1$) and color ($c$) are required, with distributions adopted from \citet{pahtheondataset}. }

{Regarding volumetric rates, CCSN subtype fractions and LF, we utilize the intrinsic SN~Ia rate from \citet{SN_RATE14} and the CCSN rate from \citet{strolger15}. Within the redshift range of this work, the volumetric rate of CCSNe is approximately five times higher than that of SNe~Ia. \citet{strolger15} applied a control-time approach to CANDELS and CLASH data to derive CCSN rates. Similarly, \citet{SN_RATE14} determined the SN~Ia rate by using theoretical delay-time distributions and observational data. The CCSN subtype fractions and LFs are adopted from \citet{V19} and \citet{J17}. Specifically, \citet{V19} provides a data-driven library of 67 spectrophotometric time-series templates covering major CCSN subtypes. For SNe~Ia, while $x_1$ and $c$ are obtained from \citet{pahtheondataset}, the absolute magnitude ($M_0$) distribution follows \citet{SH0ESMABS-19.253RIESS}. The nuisance parameters $\alpha$ and $\beta$ are taken from \citet{descollaboration2024dark}. 
Next-generation wide-field surveys, including $\it Euclid$, LSST, and CSST, will be essential for constraining CCSN rates and subtype fractions at high redshifts.}

We assume a Milky Way extinction base on \citet{f99}, with $R_V=3.1$ and $E^{\text{MW}}_{(\text{B}-\text{V})}$ around $0.01$. Here the host galaxy extinction is neglected for simplicity. More details can be found in \citet{wang2024}. We present simulated SNe based on the CSST instrumental design and the CSST-UDF survey strategy in Figure~\ref{fig:Iasnlight curve} and Figure~\ref{fig:ccsnlight curve}, including SNe~Ia at four different redshifts and six different types of CCSNe.

{In this study, since we intend to explore the capability of machine learning for SN classification in the CSST-UDF survey, we apply looser selection criteria than those used in \citet{wang2024}. These looser selection criteria will increase the total sample size, but also introduce more CCSNe. However, by utilizing the machine learning classifier, we can effectively identify these potential CCSN contaminants, obtaining a dataset with both a larger sample size and a higher purity (i.e., the fraction of true SNe~Ia within the classified SN~Ia sample).}
The selection criteria applied to our dataset are as follows:

\begin{enumerate}
    \item At least one photometric measurement with $\rm SNR>5$ before peak brightness;
    \item At least one photometric measurement with $\rm SNR>5$ after peak brightness;
    \item At least two photometric measurements with $\rm SNR>5$ in two different bands;
    \item At least three photometric measurements with $\rm SNR>5$ across all bands.
\end{enumerate}
{We generate two light curve datasets that pass the above selection criteria, i.e, a training dataset and a testing data for the machine learning. The training dataset comprises approximately 500,000 SNe, }and the testing dataset includes 9,445 SNe based on the CSST design and survey strategy. 

{Because CCSNe are generally dimmer than SNe~Ia, they often fall below the detection threshold at high redshifts. Consequently, the majority of the CCSN contamination is concentrated at intermediate redshifts ($z \sim 0.7$), as shown by \citet{wang2024}. To account for selection effects introduced by SNR and luminosity cuts, we provide a simplified bias simulation in Section~\ref{subsec:Biassim}.}

\section{Supernova Classification} \label{sec:classification}
\subsection{SuperNNova Framework} \label{subsec:snn}
Various machine learning classifiers have been developed for supernova classification {\citep{LOCHERNER_SNML2016,Charnock_2017,Villar_2021_SNML,Qu_2021SCONE,SNML_li2025machinelearningstellarastronomy,Nugent_2026_SNML,Boesky_2026_SNML}.} {The machine learning classifier is trained to learn the variations in light curves of different SN types in order to distinguish them.} In this study, we adopt the SNN \citep{SuperNNova} as the main framework. SNN is based on RNNs, which are widely used for analyzing time-series data. The framework includes several built-in methods, such as Random Forest (RF), Long Short-Term Memory (LSTM) networks, and Bayesian Neural Networks (BNNs). 
We use the LSTM-based model in this work, {as its architecture can help reduce the vanishing and exploding gradient issues and provides more stable and reliable performance \citep{hochreiter1997LSTM,chung2014_LSTM,Charnock_2017}.}

\begin{figure}[htbp]
    \centering
    \includegraphics[width=0.47\textwidth]{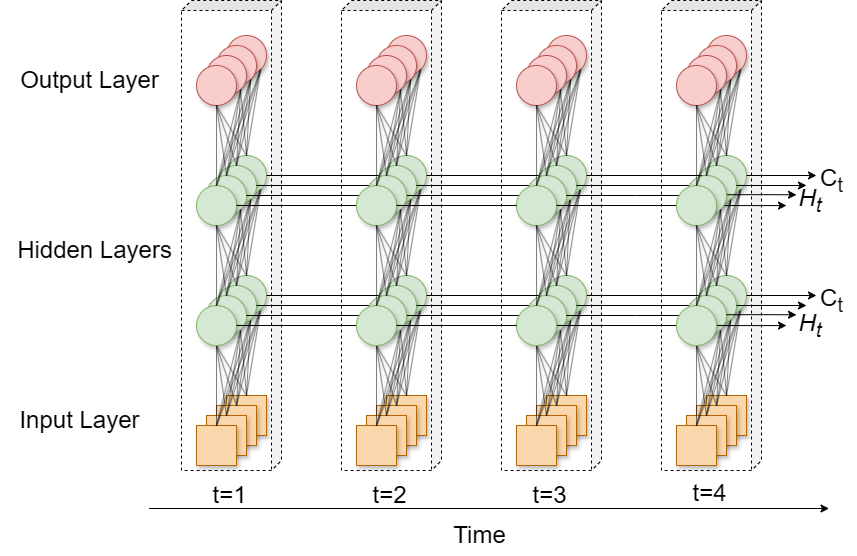} 
    \caption{A simplified architecture of an LSTM network. The cell state (\(C_t\)) carries information through time. The input layer receives data at each time step, while hidden layers compute and update the hidden state (\(H_t\)). The output layer generates predictions. The diagram illustrates the flow of information across time steps \(t \in [1, 4]\). }
    \label{fig:lstmstruct}
\end{figure}

\begin{figure}[htbp]
    \includegraphics[width=0.47\textwidth]{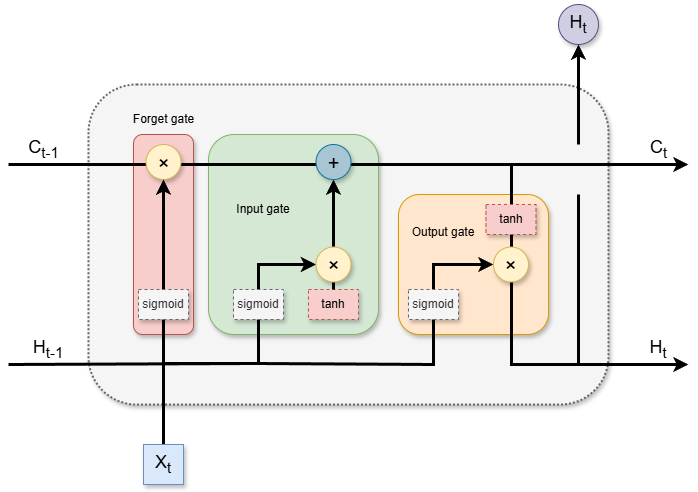}
    \caption{An LSTM cell within the hidden layer, comprising the cell state (\(C_t\)), hidden state (\(H_t\)), and input (\(X_t\)). It includes key components: Input Gate, Forget Gate, and Output Gate, which enable the model to effectively capture long-term dependencies.}
    \label{fig:lstmcell}
\end{figure}

In Figure~\ref{fig:lstmstruct}, we illustrate a simplified architecture of an LSTM network and its information flow across time steps $t \in [1, 4]$. The cell state ($C_t$) carries information through time steps, while the input layer processes sequential data. Hidden layers update the hidden state ($H_t$), and the output layer generates predictions. 

Compared to a basic RNN, an LSTM network improves performance by introducing three distinct gates: input, forget, and output as shown in Figure~\ref{fig:lstmcell}. These gates, together with a dedicated cell state, act as internal memory that allows the network to decide what information to keep or discard.
The input and forget gates use sigmoid activations to regulate the cell state, while the output gate controls how much of the processed information is passed on to the hidden state $H_t$. These mechanisms allow the network to handle long-term dependencies effectively and avoid  gradient problems in RNNs.
Because of these properties, LSTM networks are particularly suited for modeling the complex temporal evolution of SN light curves, making them highly effective for classification tasks \citep{chung2014_LSTM,Charnock_2017}.
Furthermore, the SNN provides a flexible framework with multiple training options, such as the choice of network architecture, number of layers, learning rate, normalization method, input redshift type (none, photometric, or spectroscopic), and number of classification categories. 

{The inputs to the model include the relative observing time, the photometric fluxes, and the corresponding flux errors across all available filter bands, as well as the host galaxy redshift (if available).} The model outputs the probability that a given SN belongs to a specific SN subtype and can also provide early-phase classification results as the light curve evolves. SNN has already been successfully applied to real survey data \citep{DES5Year23,des5yrNhz}, {showing high classification precision and strong robustness under various observing conditions.
In this study, we focus on classification results obtained using the full light curve data, up to a phase of +30 days}
\subsection{Model Training}\label{subsec:modeltrain}
The generation of the training dataset is performed using the {\tt SNCosmo} framework, based on the CSST-UDF survey strategy, SN parameters, and light curve selection criteria described in Section~\ref{sec:sn-sim}. After balancing the sample sizes, the final training dataset comprised approximately 250,000 SNe~Ia and 250,000 CCSNe, We ensure that this balance does not impact the distribution of various SN parameters. {Although this 1:1 ratio does not reflect the intrinsic physical rates, it is explicitly adopted to ensure that the model assigns equal weight to learning the decision boundaries of both classes during training. This approach prevents the classifier from being biased toward the majority class \citep{hochreiter1997LSTM,LOCHERNER_SNML2016,SuperNNova}.}

Subsequently, we train the classification model using the SNN with a two-layer LSTM network. Each LSTM layer contains 32 hidden units in each direction, resulting in 64 features after bidirectional concatenation. A dropout rate of 0.05 is applied between layers to prevent overfitting, and the normalization scheme of SNN is set to ''cosmo''. In the model training process, we do not include host galaxy redshift as an input feature for conservation purpose, although CSST spectroscopic UDF survey may measure the redshifts of a large fraction of host galaxies \citep{wang2024}. We also test the models with more layers and hidden units, but we find that the results do not show significant improvement. To avoid potential issues such as overfitting and unnecessary model complexity, we keep the current network parameters.

\subsection{Classification} \label{subsec:classres}
We generate a testing dataset containing 9,445 SNe based on the CSST-UDF survey strategy, SN parameters, and the same selection criteria as the training dataset as described in Section~\ref{sec:sn-sim}, which includes the observational fluxes, times, filters, and SN identifiers. {This testing dataset completely follows the realistic physical proportions and the CSST selection effects of SNe Ia and CCSNe as described in Section~\ref{sec:sn-sim}. The trained SNN model is then applied to classify this testing sample.}

\begin{table}[t]
\centering
\caption{Binary confusion matrix of the SN classification on the testing dataset. $\it TP$, $\it TN$, $\it FP$, $\it FN$, and $\it Total$ denote the true positive, true negative, false positive, false negative, and total samples, respectively.}

\label{tab:confusion_matrix}
\renewcommand{\arraystretch}{1.4} 
\begin{tabular}{l | c c | c}
\hline
 & \multicolumn{2}{c|}{\textbf{Predicted Class}} & \\
\cline{2-3}
\textbf{True Class} & \textbf{SN Ia} & \textbf{CCSN} & {\it Total} \\
\hline
\textbf{SN Ia} & \textbf{2864} ($\it TP$) & 11 ($\it FN$) & 2875 \\
\textbf{CCSN}    & 74 ($\it FP$) & \textbf{6496} ($\it TN$) & 6570 \\
\hline
{\it Total}   & 2938 & 6507 & \textbf{9445} \\
\hline
\end{tabular}
\end{table}

{In Table~\ref{tab:confusion_matrix}, we show the binary confusion matrix which summarizes the exact distribution of the classification results. We can find that, for the 9,445 testing SNe, the classifier can achieve an overall accuracy of 99.1\%, which is defined as the fraction of all correct predictions of SN~Ia and CCSN, i.e. $(TP+TN)/Total$. Applying a probability cut of $P_{\text{Ia}} > 50\%$, it identifies 2,938 SN~Ia candidates (2,864 true SNe~Ia and 74 CCSN contaminants), yielding an initial purity of 97.5\%, which is defined as $TP / (TP + FP)$. Specifically, the 74 false positive samples consist of 14 Type II SNe and 60 Type Ib/c SNe (35 Type Ib and 25 Type Ic). This explicitly indicates that Type Ib/c SNe are more difficult to isolate, since their relatively bright absolute magnitudes and light-curve shapes are similar to those of SNe~Ia.}



%

\begin{figure}[h]
    \centering
    \includegraphics[width=1\linewidth]{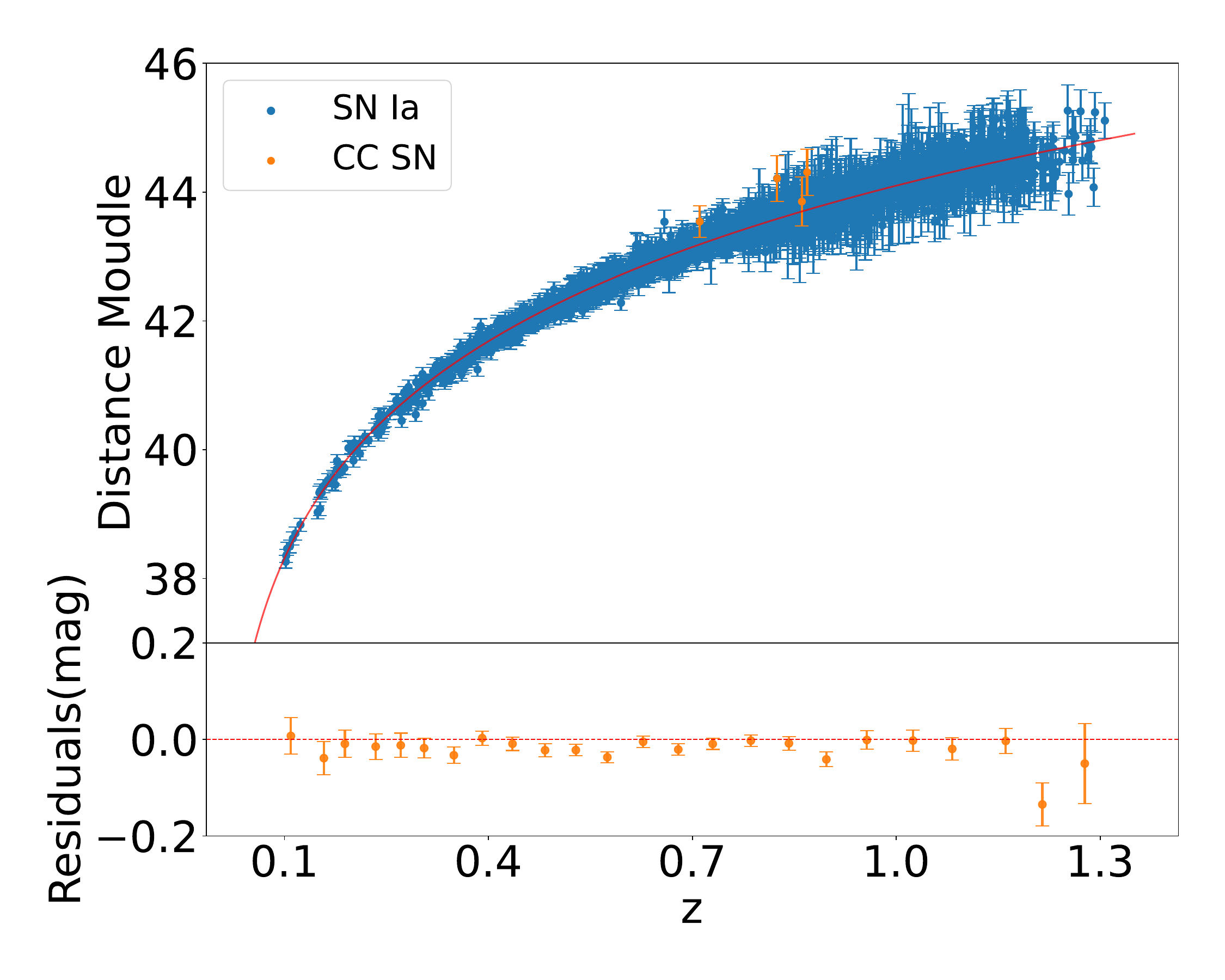}
    \caption{{Hubble diagram as a function of observed redshift for the 2197 SNe classified by SNN. Blue and orange points represent the observed distance module of SNe~Ia and CCSNe, respectively. The red solid line indicates the fiducial cosmological model. The lower panel displays the distance modulus residuals and associated errors relative to this fiducial cosmology across 24 redshift bins.}}
    \label{fig:HD}
\end{figure}

Next, following the fitting procedures in \citet{wang2024}, we perform SALT3 template fitting on these candidates to derive the SN~Ia light curve parameters, including redshift $z$, time of maximum brightness $t_0$, amplitude $x_0$ (or $m_B$), stretch $x_1$, and color $c$. Then we apply a relaxed version of the Joint Light-curve Analysis like (JLA-like) criteria for further selection, which are defined as 
\begin{enumerate}
    \item Reduced chi-square \(\chi^2_{\text{Reduced}} > 5\).
    \item Best-fit values of \(x_1\) or \(c\) outside the SALT3 model valid ranges: \(x_1 \notin (-3, 3)\) or \(c \notin (-0.3, 0.3)\).
    \item Error on the time of peak brightness exceeding 2: \(t_{\text{0,err}} > 2\).
\end{enumerate}
Finally, we obtain a high-purity photometric SN~Ia sample exceeding 99.5\%. As shown in Figure~\ref{fig:HD}, this sample contains 2,197 SNe with only 4 remaining CCSN contaminants. {Compared to traditional selection methods \citep{wang2024}, our approach increases the total sample yield of SNe~Ia for cosmological analysis by $\sim$10\% without sacrificing purity.}

We should also note that, in the real observations, there can be greater challenges in the SN classification, which could result in an lower purity less than $98\%$ \citep[e.g.][]{DES_BIASFROMCLASS}, considering the systematic uncertainties such as dust extinction, color steps, photometry error and calibration error \citep{DES_SYS_vincenzi2024}.
In the future, we will investigate these effects in more details and assess the impacts on the SN classification.


\section{Cosmological constraint} \label{sec:cosmology}

In the cosmological constraint only using the SN photometric data, the dataset with CCSN contamination can be directly utilized. In this kind of method, a simplified CCSN likelihood is assumed, including the constant offset, linear offset, and quadratic offset for SN~Ia deviations, and the cosmological and nuisance parameters, e.g. $\alpha$, $\beta$, the absolute magnitude $M_0$, the contamination rate, etc., are jointly fitted \citep[e.g.][]{gong2010,wang2024}. {Our tests reveal that at low contamination levels (as shown in Figure~\ref{fig:HD}), using a global contamination rate may fail to recover the fiducial cosmological parameters, and introducing too many free parameters can significantly increase the constraint uncertainties.}


Here we employ SNN to derive SN classification probabilities, and the cosmological parameters are then determined using a two-step binned analysis, applying the Bayesian Estimation Applied to Multiple Species (BEAMS) with Bias Corrections (BBC) framework to correct the SN~Ia magnitude bias caused by the selection effect and CCSN contamination. Note that we mainly consider the selection effect here, since the remaining CCSN contamination is very low and can be neglected in our case.

\subsection{Selection Effect}\label{subsec:Biassim}
The selection effect presents a significant challenge in SN~Ia observations and statistical inference. To assess this influence for the CSST-UDF survey, we perform a simplified simulation to estimate the average magnitude bias of SNe in different redshift bins.
{Specifically, we conduct 50 sets of selection-effect simulations using the exact configuration described in Section~\ref{sec:sn-sim}. To increase the total number of SNe while preserving their original parameter distributions, we enlarge the survey area by a factor of five instead of increasing the survey time.}
Each simulation follows four sequential steps. First, we generate mock catalogs under the CSST-UDF observing conditions and apply selection criteria described in Section~\ref{sec:sn-sim}, leaving roughly 11,000 SNe~Ia per set. Next,we calculate the Hubble residuals as { 
$\Delta \mu_i = \mu_{{\rm obs},i} - \mu_{\rm fid}(z_i,\vec\theta_{\rm fid}),$}
where
\begin{equation}
\mu_{{\rm obs},i} = m^{\rm fid}_{B,i} + \alpha x^{\rm fid}_{1,i} - \beta c^{\rm fid}_i + M^{\rm fid}_0,
\end{equation}
$m^{\rm fid}_{B,i}$, $x^{\rm fid}_{1,i}$, $c^{\rm fid}_i$, and $M^{\rm fid}_0$ {are the individual fiducial parameters of each SN.
The fiducial distance modulus of each SN, $\mu_{\rm fid}(z_i,\vec\theta_{\rm fid})$, is computed from the fiducial cosmological parameter vector $\vec\theta_{\rm fid}$ and is given by }
\begin{equation}
\begin{aligned}
\mu_{\rm fid}(z) &= 5 \log_{10} d_{\rm L}(z) + 25,\\
d_{\rm L}(z)&= \frac{(1+z)}{H_0} \int_0^z \frac{c\,{\rm d}z^{\prime}}{E(z)},
\end{aligned}
\label{eq:DL}
\end{equation}
where the reduced Hubble parameter $E(z)$ is calculated as
\begin{equation}
E(z) =\left[\Omega_{\rm DE}(1+z)^{3(1+w)}+\Omega_{\mathrm{M}}(1+z)^3 \right]^{1/2}.
\label{eq:EZ}
\end{equation}
Here $H_0$ is the Hubble constant and $w$ is the dark energy equation of state parameter. $\Omega_{\rm DE}$ and $\Omega_{\rm M}$ are the energy density parameters of dark energy and dark matter, and $\Omega_{\rm DE}=1-\Omega_{\rm M}$ in a flat universe.

Then, we divide SNe in the range $0.1 < z < 1.3$ into 24 equally spaced redshift bins. In each simulation and within each bin, we fit the Hubble residuals of all SNe in that bin with a Gaussian distribution, and take the mean of the Gaussian as the bias for that bin. This process is repeated across all 50 simulations, producing 50 independent bias estimates per bin. Finally, the 50 bias values in each bin are themselves fitted with a Gaussian function, and the mean of this distribution represents the average magnitude bias uncertainty across simulations.

Our analysis indicates that magnitude bias caused by the selection effect becomes increasingly significant at higher redshifts, with values of $0.014\,\mathrm{mag}$, $0.021\,\mathrm{mag}$, $0.028\,\mathrm{mag}$, and $0.036\,\mathrm{mag}$ for the redshift bins centered at $z = 1.125$, $1.175$, $1.225$, and $1.275$, respectively. {For comparison, the intrinsic dispersion in the absolute magnitude of SNe~Ia is about $0.1\,\mathrm{mag}$ \citep{SH0ESMABS-19.253RIESS,2021Brout_sigma0.1}}. These simulations provide a quantitative measure of how selection effect influences supernova surveys.
{Therefore, to reduce the SN~Ia magnitude bias and the systematic bias in the cosmological constraint, the BBC method is employed in our analysis.}

\subsection{BBC method}\label{subsec:BBC}
 
The Bayesian Estimation Applied to Multiple Species (BEAMS) with Bias Corrections (BBC) method provides an effective way to correct the SN~Ia magnitude bias due to the selection effect and CCSN contamination. Accordingly, we adopt the two-step binning BBC method~\citep{M11,descollaboration2024dark}. The BBC binning method is divided into two steps: first, the SNe are assigned into different redshift bins, and the offsets of SNe relative to the reference cosmology within each bin are fitted. In the second step, the reference cosmology and these offsets are used to constrain the cosmological parameters.

Several studies have shown that the influence of the assumed reference cosmology on cosmological constraints becomes negligible when a sufficiently large number of bins is used \citep{M11,descollaboration2024dark,DES_2024beyondLCDM}. In our analysis, the dataset spans a redshift range $0.1 \leq z \leq 1.3$, and we divide it into 24 redshift bins that are equally spaced in log-space. Some analyses, e.g. DES, use an equal number of SNe per bin \citep{descollaboration2024dark}, and our tests show that the binning method has a minimal impact on the results.

The first step of the BBC method fits the $\Delta\mu$ offsets across redshift bins using binned SNe and a set of reference cosmological parameters. Following the formalism of \cite{BBCNOBIN2023K}, the simplified BBC likelihood per event is given by
\begin{equation}
\mathcal{L}_i \propto P_{\mathrm{Ia},i  }D_{\mathrm{Ia},i} + (1-P_{\mathrm{Ia},i})D_{\mathrm{CC},i},
\end{equation}
where $D_{\mathrm{CC},i}$ describes the CCSNe term of the total likelihood function, which is generally modeled by an assumed function or derived from simulations~\citep{BEAMSH12,KS17BBC,DES_BIASFROMCLASS}. In this work, since the number of CCSNe is very small after the RNN classification, whose effect can be neglected, this term is not included in the analysis. $P_{\mathrm{Ia},i}$ denotes the photometric classification probability that the $i$-th SN is a SN~Ia, and the core components are
\begin{equation}
D_{\mathrm{Ia},i} = \exp\left[-\chi_{\mathrm{HR},i}^2/2\right], \quad
\chi_{\mathrm{HR},i}^2 = \frac{\mathrm{HR}_i^2}{\sigma_i^2}.
\label{equ:HR}
\end{equation}
The Hubble residual $\mathrm{HR}_i$ for a SN~Ia is defined as
\begin{equation}
\begin{aligned}
\mathrm{HR}_i &= \mu_i - \left[\mu_{\mathrm{ref}}(z_i,\vec{\theta}_{\mathrm{ref}}) + \Delta_{\mu,\zeta}\right],\\
\mu_{\mathrm{ref}}(z_i,\vec{\theta}_{\mathrm{ref}}) &= 5 \log_{10}\left[(1+z) \int_0^z \frac{\,{\rm d}z^{\prime}}{E(z)}\right]+ 25,
\end{aligned}
\label{equ:HRi}
\end{equation}
where $E(z)$ follows Equation~(\ref{eq:EZ}), $\Delta_{\mu,\zeta}$ is the distance modulus offset within the $\zeta$-th redshift bin, which serves as a free parameter in the first step of fitting. Following the determination of $\Delta_{\mu,\zeta}$, the second step of fitting is subsequently performed. $\mu_i$ is the observed distance modulus, defined as
\begin{equation}
\mu_i=m_{B,i}+\alpha x_{1,i}-\beta c_i+\mathcal{M},
\end{equation}
where $\alpha$ and $\beta$ are the stretch and color-luminosity parameters, treated as nuisance parameters in the fitting process. The values of $m_{B,i}$, $x_{1,i}$, and $c_i$ are the SALT3 parameters estimated from the light curve fitting.
$\mathcal{M}$ is a combination of the absolute magnitude $M_0$ and the Hubble constant $H_0$ , which can be expressed as
\begin{equation}
\mathcal{M}=M_0+5\log_{10}(c/H_0).
\end{equation} 
This parametrization of $\mathcal{M}$ decouples $H_0$ from other cosmological parameters, ensuring that the inferred value of $H_0$ does not impact the fitting of the remaining cosmological parameters.
$\mu_{\mathrm{ref}}(z_i,\vec{\theta}_{\mathrm{ref}})$ in Equation~(\ref{equ:HRi}) represents a reference distance modulus, typically calculated using either a cosmological model with parameter vector $\vec{\theta}_{\mathrm{ref}}$ or polynomial redshift functions. When the number of bins is sufficiently large, we notice that the choice of reference cosmology has negligible impact on the results~\citep{M11}.

The total uncertainty $\sigma_i$ in Equation~(\ref{equ:HR}) includes several components, which is given by
\begin{equation}
\sigma_i^2 = \sigma_{\mathrm{int}}^2 + \sigma^2_{\mu,z} + \sigma^2_{\mu,i},
\end{equation}
{where $\sigma_{\mathrm{int}}=0.1$ \citep{SH0ESRiess_2019} }is the SN~Ia intrinsic uncertainty, $\sigma_{\mu,z} = \frac{5}{\ln(10)} \frac{1+z}{z(1+z/2)} \sigma_z$ quantifies the uncertainty arising from photometric redshift errors, and $\sigma_{\mu,i}^2$ derives from the light curve fitting covariances of $m_B$, $x_1$, and $c$.

We adopt a simplified version of the BBC likelihood, and it can be written as
\begin{equation}
\mathcal{L}_{\mathrm{total}} \propto \prod_{i=1}^N \mathcal{L}_i \propto \prod_{i=1}^N P_{\mathrm{Ia},i}D_{\mathrm{Ia},i}.
\label{equ:LH}
\end{equation}
Here, we only consider the SN~Ia part, since as described in Section~\ref{subsec:classres}, the high {classification performance} of SNN ensures an extremely low CCSN contamination rate in the CSST-UDF SN dataset, with minimal impact from residual contaminants. 

We employ the {\tt emcee} package \citep{emcee} using the Markov Chain Monte Carlo (MCMC) method in the fitting process. The known parameters are $m_{B,i}$, $x_{1,i}$, $c_i$, and the reference cosmological parameter vector $\vec{\theta}_{\mathrm{ref}}$. The average absolute magnitude of the SN~Ia is fixed at $M_0=-19.25$. We test two reference cosmological parameter sets: ($\Omega_{\rm M}=0.28$, $w=-1.3$, $H_0=72$) and ($\Omega_{\rm M}=0.35$, $w=-1.3$, $H_0=73$). Our results show that the choice of reference cosmology has a negligible impact on the final parameter estimates. The model parameters to be fitted in the MCMC are $\alpha$, $\beta$, and the $\Delta_{\mu,\zeta}$ offsets in 24 different redshift bins. We set the parameter ranges as follows: $\alpha\in(0.08, 0.32)$, $\beta\in(1, 5)$, and $\Delta_\mu\in(-5, 5)$. We generate 200 chains, each contains 10,000 points after burn-in, and 40,000 points are retained after the thinning process.

After completing the first step BBC fitting, we obtain the distance modulus offsets $\Delta_{\mu,\zeta}$ and $\sigma_{\Delta_{\mu,\zeta}}$ in 24 redshift bins along with $\alpha$ and $\beta$. Based on these results, we proceed to the second step BBC fitting, where the sum of the reference distance modulus $\mu_{\mathrm{ref}}$ and the offsets $\Delta_{\mu,\zeta}$ collectively describe the cosmological distance-redshift relation.

The second step of the BBC fitting procedure uses the reference cosmological parameters, redshift bin offsets, and their uncertainties to constrain the cosmological parameters. The best-fit values of the cosmological parameters are obtained by maximizing the likelihood function 
\begin{equation}
\begin{split}
\mathcal{L} \propto \exp\left[-\chi^2_\Delta / 2\right], \  \  \  
\chi^2_\Delta = \mathbf{D}_{\mu,\zeta}^\top \mathcal{C}^{-1} \mathbf{D}_{\mu,\zeta}, \\
\mathbf{D}_{\mu,\zeta} \equiv \Delta_{\mu,\zeta} + \mu_{\mathrm{ref},\zeta} - \mu(\vec{\theta},z_\zeta ),
\label{eq:LH2}
\end{split}
\end{equation} 
where $\Delta_{\mu,\zeta}$, $z_\zeta$, and $\mu_{\mathrm{ref},\zeta}$ denote the BBC-fitted distance modulus offset, effective redshift, and  reference distance modulus in redshift bin $\zeta$, respectively. $\vec{\theta}$ is the parameter vector to be fitted in the MCMC. The covariance matrix $\mathcal{C}$ is constructed directly from the uncertainties of $\Delta_{\mu,\zeta}$, i.e. $\sigma_{\Delta_{\mu,\zeta}}$, including both statistical and systematic contributions, with only the diagonal elements retained. 
The effective redshift $z_\zeta$ of each bin is calculated from the inverse distance modulus function:
\begin{equation}
{z_\zeta} = \mu^{-1}(\overline{\mu_{\mathrm{ref,\zeta}}}), 
\end{equation}
where the value of $\overline{\mu_{\mathrm{ref,\zeta}}}$ is the weighted average of the reference distance moduli within bin $\zeta$, with weights given by $\sigma_\mu^{-2}$ for each supernova.

In the second step of the BBC fitting, we perform the MCMC analysis using the results obtained from the first step. The known quantities are the reference cosmological parameters, and the $\Delta_{\mu,\zeta}$ data and its uncertainty $\sigma_{\Delta_{\mu,\zeta}}$, as shown in the lower panel in Figure \ref{fig:HD}. The model parameters $\vec{\theta}$ to be fitted are $\Omega_{\rm M}$ and $w$ assuming the flat $w$CDM model, with priors $\Omega_{\rm M} \in (0, 0.5)$ and $w \in (-2, 0)$. We run 100 MCMC chains, each consists of 10,000 steps after burn-in, and 20,000 points are retained after the thinning process.

\begin{figure}
    \centering
    \includegraphics[width=1\linewidth]{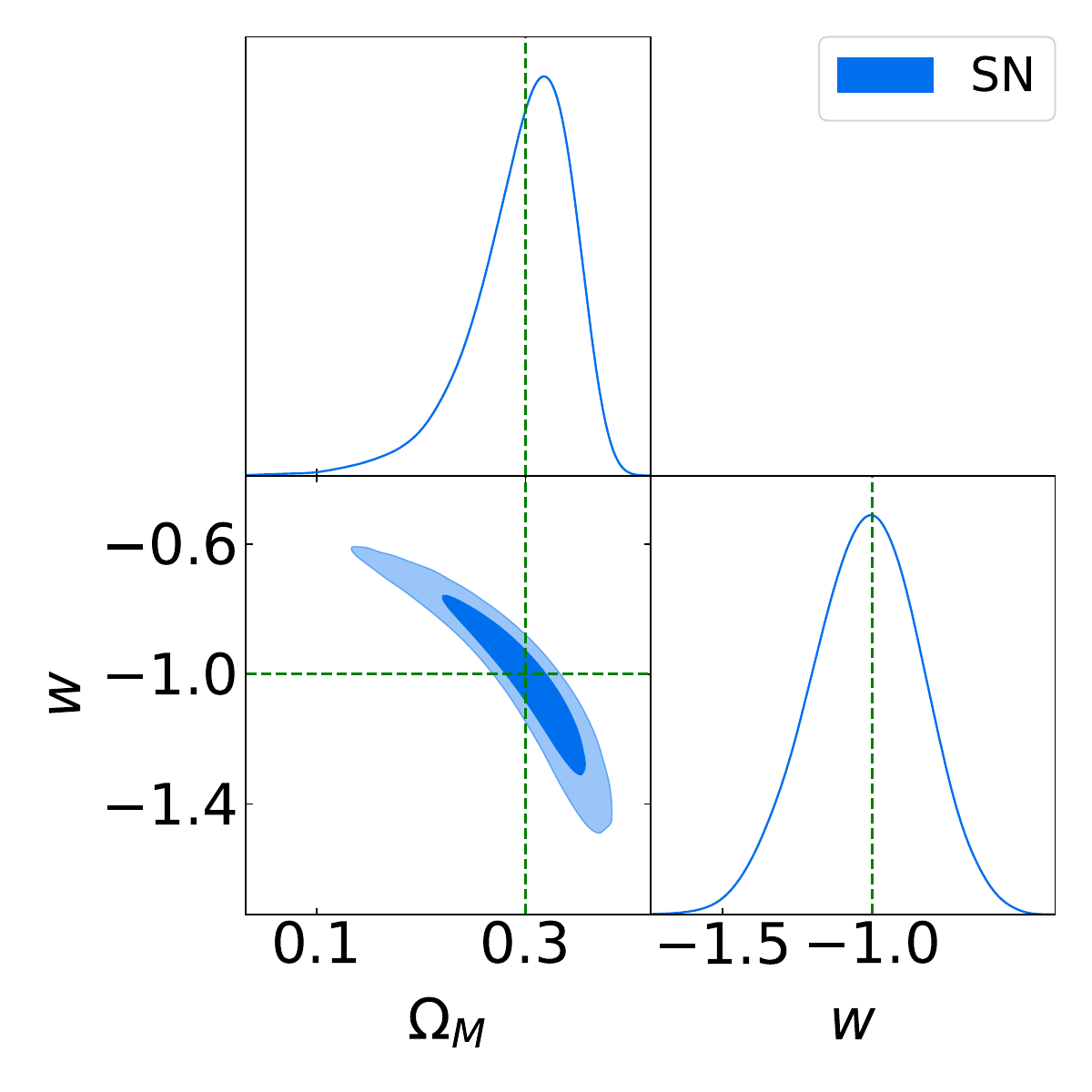}
    \caption{{Predicted 1$\sigma$ and 2$\sigma$ confidence contours and 1-D PDFs of $\Omega_{\rm M}$ and $w$ for the flat $w$CDM model in the CSST-UDF SN photometric survey.  The green dashed lines indicate the input fiducial cosmological parameters used in the simulation.}}
    \label{fig:Flat_Cosmo_Result}
\end{figure}

For the classified CSST-UDF SN photometric mock dataset, we find that $\Omega_{\rm M} = {0.304}^{+0.030}_{-0.052}$ and $w = {-1.017}^{+0.177}_{-0.189}$, corresponding to the $1\sigma$ relative accuracies of 14\% for $\Omega_{\rm M}$ and 18\% for $w$. In Figure~\ref{fig:Flat_Cosmo_Result}, we show the 1D marginalized probability distribution functions (PDFs) and 2D contour maps (1$\sigma$ and 2$\sigma$) of $\Omega_{\rm M}$ vs. $w$. These results indicate that, despite relying solely on photometrically classified SNe, our method achieves constraint precision comparable to those obtained from spectroscopically confirmed samples in previous surveys~\citep[e.g.][]{PANTHEONCOSMOBrout_2022}. 

\section{Summary}\label{sec:conclusion}
In this study, we generate the mock light curve data of SN~Ia and CCSN in the CSST-UDF photometric survey, and investigate the SN classification with the RNN machine learning method by applying the SNN framework. Compared to traditional photometric classification techniques, the SNN, in conjunction with the JLA-like cuts, can achieve a significant improvement in classification purity, exceeding 99.5\% when the full light curve data are utilized. This has greatly reduced the contamination from CCSNe, which is essential for the unbiased cosmological constraints. Furthermore, we optimize the cosmological constraint workflow by incorporating the BBC-like methods. This adjustment allows the cosmological analysis to better reflect realistic survey conditions and yields more reliable constraints. 

We find that the cosmological constraints on $\Omega_{\rm M}$ and $w$ can achieve the accuracies of 14\% and 18\%, respectively, assuming the flat $w$CDM model. These results are comparable to those from the surveys that relied on spectroscopic confirmation, demonstrating the potential of the SN~Ia cosmology with photometric data only in the next generation surveys such as CSST.

\begin{acknowledgments}
M.L.W. and Y.G. acknowledge the support from the CAS Project for Young Scientists in Basic Research (No. YSBR-092), and the National Key R\&D Program of China grant Nos. 2022YFF0503404 and 2020SKA0110402. X.L.C. acknowledges the support of the National Natural Science Foundation of China through grant Nos. 11473044 and 11973047 and the Chinese Academy of Science grants ZDKYYQ20200008, QYZDJ- SSW-SLH017, XDB 23040100, and XDA15020200. This work is also supported by science research grants from the China Manned Space Project with grant Nos. CMS-CSST-2025-A02, CMS-CSST-2021-B01, and CMS-CSST-2021-A01.
\end{acknowledgments}

%

\vspace{5mm}


\software{
          SuperNNova \citep{SuperNNova}, 
          SNCosmo \citep{SNCOSMO},
          Astropy \citep{ASTROPY},  
          Emcee \citep{emcee}. {The code and data that support the findings of this study are available from the corresponding author upon reasonable request.}
          }


%




\bibliography{AA_CITE}{}
\bibliographystyle{aasjournal}



\end{document}